# All Magnesium diboride Josephson Junctions with MgO and native oxide barriers

M. V. Costache [a)] and J. S. Moodera

*Francis Bitter Magnet Laboratory, Massachusetts Institute of Technology*

*Cambridge, MA 02139*

We present results on *all*-MgB$_2$ tunnel junctions, where the tunnel barrier is deposited MgO or native-oxide of base electrode. For the junctions with MgO, the hysteretic I-V curve resembles a conventional underdamped Josephson Junction characteristic with critical current-resistance product nearly independent of the junction area. The dependence of the critical current with temperature up to 20 K agrees with the Ambegaokar-Baratoff expression. For the junctions with native-oxide, conductance at low bias exhibits sub-gap features while at high bias reveals thick barriers. As a result no supercurrent was observed in the latter, despite the presence of superconducting-gaps to over 30 K.

[a)] E-mail address: costache@mit.edu

Magnesium diboride (MgB$_2$), a compound known since the 1950s, was recently[1] discovered to be superconductor at a remarkably high critical temperature T$_C$ = 40 K. Its high T$_C$, simple crystal structure, large coherence lengths, and high critical current densities and fields indicate that MgB$_2$ has potential for superconducting devices that operate at 20-30 K, the temperature reached by current commercial cryocoolers.[2,3] Further, its large energy gap of about 6-8 meV, will potentially allow higher operation speeds than Nb-based superconductor devices. To make use of this great potential, however, one needs to develop a standard process to fabricate *all* MgB$_2$ Josephson Junctions. This should in theory be relatively straightforward, but under practical conditions huge challenges appear mainly due to difficulty in creating a leakage-free ultra thin tunnel barrier that can also survive high-temperature (typically $\geq$ 300 °C) deposition of the top MgB$_2$ electrode.

Study of *all* MgB$_2$ Josephson Junctions has been reported by several groups, but to date, the reproducibility of the results has been low. *All* MgB$_2$ junctions were obtained by Shimakage *et al*. with AlN,[4] Ueda *et al*.[5] and Shim *et al*.[6] with an Al$_2$O$_3$ as the barrier material. However, due to poor reproducibility of the results, caused by the presence of some form of natural oxide at the interfaces, the Al-based barriers seem not reliable. In addition, due to high-temperature deposition of the top MgB$_2$ electrode the unoxidized Al metal can diffuse into MgB$_2$ film and it degrades superconductor characteristics.[6] Driven by the difficulty to form a good artificial barrier, recent studies have focused on forming the native oxide barrier obtained by oxidation of base MgB$_2$ electrode.[7-9] However, the process that forms this tunnel barrier and its composition is not yet understood. Some unreacted Mg and B that likely remain at the surface of base MgB$_2$ film will react during the fabrication of tunnel barrier. Nevertheless, MgB$_2$/oxide/Pb Josephson Junctions fabricated by Cui *et al*.[9] using hybrid physical-chemical vapor deposition showed good Josephson characteristics, suggesting that the natural oxide may be suitable for *all* MgB$_2$ Josephson Junctions technology. Another material is MgO, well-proven to survive high temperature fabrication process in magnetic memories technology and it is also one of the earliest barrier materials used by Mijatovic *et al.*[10] in *all* MgB$_2$ multilayer ramp type junctions.

In this letter, we report fabrication of *all* MgB$_2$ junctions with a good coverage of the bottom electrode by the MgO barriers, demonstrated by a nearly constant junction resistance (R$_N$) area product. Further the Josephson characteristic voltage (I$_C$R$_N$) ranges from 2 to 2.4 mV that are the highest values reported to date. In addition, we have fabricated MgB$_2$/native oxide/MgB$_2$ tunnel junction, where the conductance at high bias reveals thick tunnel junctions and consequently no possibility for pair tunneling.

We fabricated *all* MgB$_2$ tunnel junctions with MgO barriers using photolithography and etching techniques. The films were deposited on Si (111) substrates in a molecular-beam epitaxy system with a base pressure of $10^{-10}$ Torr. Mg from a K-cell and B using e-beam were coevaporated onto Si with a MgO buffer layer at 290 °C. After cooling down to 150 °C, MgO tunnel barrier was formed by e-beam evaporation from MgO crystal pieces. Subsequently, the substrate temperature was raised to 290 °C for the top $MgB_2$ film deposition and the growth condition of the top $MgB_2$ film was identical with the bottom film. Further upon cooling down to room temperature, the top $MgB_2$ film was capped with a 4 nm thin gold layer to prevent oxidization. No post annealing was done. The final structure was Si/MgO(5 nm)/ $MgB_2$ (30 nm)/ MgO/ $MgB_2$ (30 nm)/Au, where MgO thickness was varied from 2 to 4 nm. Following this the layer structure was patterned to create junctions of area 20×20, 40×40, 60×60 and 80×80 $\mu m^2$, by photolithography and Ar ion milling. For the junctions described in this work the Josephson penetration depth $\lambda_J$ = sqrt($\Phi_0/2\pi\mu_0 dJ_C$) is in the order of 100 μm, where $d$ is the sum of barrier and $MgB_2$ thickness, $\Phi_0$ is the magnetic flux quantum and $J_C$ is the junctions critical current density (we used 10 A/$cm^2$). Since the width of the junctions was smaller than $4\lambda_J$, the junctions are in the small junction limit. As a result, a uniform current distribution due to self-field effects is expected.

In addition, using shadow mask technique we have fabricated $MgB_2$/native oxide/$MgB_2$ tunnel junctions where the barrier was prepared by oxidation of base $MgB_2$ electrode. After deposition of base $MgB_2$ electrode the substrate was cooled to room temperature, then this layer was plasma oxidized in the load lock chamber for 30 sec at 65 mTorr oxygen pressure. In this case, the junction area was 100×150 $\mu m^2$ and resistance ranges from 1 to 10 kΩ. The electrical transport measurements of the junctions were performed in a four-point configuration in a partially-shielded cryostat, in a temperature range from 1 to 40 K. The $MgB_2$ film resistivity at room temperature was ρ = 25 μΩ-cm, while residual resistance ratio $\rho(300 K)/\rho(T_C) \sim$ 2-3. The dependence of the $T_C$ on the $MgB_2$ film thickness is shown in Fig. 1a. In the context of our discussion it is important to note that the bottom electrode had a $T_C$ of 29-30 K, while top electrode had few degrees lower likely due to the less perfect film growth.[6,8]

Figure 1b shows a typical current versus voltage (I-V) measurement of $MgB_2$/MgO/$MgB_2$ junctions at 4.2 K, for a junction with area of 80×80 $\mu m^2$ and MgO thickness of 2.4 nm. We find that the 2.4 nm is the optimal thickness for our junctions, below this value the I-V data shows sub-gap conductance whereas for higher values, the transport is dominated by quasiparticle current. This is directly related to the bottom film roughness. The hysteretic I-V curve resembles a conventional underdamped junction characteristic, using resistively and capacitively shunted junction model the Stewart-McCumber parameter was found to be 12. As expected the $I_CR_N$ product is nearly independent of the junction area (Fig. 1c). At 4.2 K the $I_CR_N$ magnitude ranges from 2 to 2.4 mV, smaller than theoretical predicted 4 mV[13] whereas they are highest among values reported as yet.[4-6,9] The small variation of the $I_CR_N$ product with junction area implies a good uniformity of the MgO barrier. The temperature dependence of $I_C$ is shown in Fig. 2a. As seen the junctions show $I_C$ up to 21 K. The $I_C$ dependence on temperature was fitted with Ambegaokar–Baratoff expression[11] $I_C(T) = (\pi \Delta(T)/2eR_N)\tanh(\Delta(T)/2k_BT)$ with a BCS-like temperature dependence of the gap, shown in Fig. 2a. The best fit was obtained by taking $T_C$'s 13 and 21 K, for the two superconductor electrodes. As already mentioned the top MgB$_2$ layer has a lower $T_C$ due to poorer crystallinity and may have degraded further during the patterning process.[12] However, we expect its $T_C$ could not be as low as the fit showed. Thus we believe that the thermal noise suppressed $I_C$ the effect at higher temperatures. At 4.2 K, the Josephson coupling energy ($E_J=h\Delta/8e^2R_N \sim 30$ meV) is greater that the thermal energy ($k_BT \sim 0.36$ meV), but at 20 K the Josephson energy is only few times higher than the thermal energy. Hence, it is evident that transport is seriously influenced by the thermal noise. In addition, it is important to mention that in all experiments where Josephson effect was observed so far[4-6] the $I_C$ disappeared below or near 20 K. In this instance, it is necessary to further study whether the lack of supercurrent above 20 K in *all* MgB$_2$ tunnel junctions is due to fabrication process or thermal noise or some characteristic of multiband superconductivity as previously noted.[3,13]

The ac Josephson effect was measured for the junctions at 9.3 GHz. Figure 2b shows the I-V characteristics of a 60 x 60 µm$^2$ junction at 4.2 K as a function of irradiated power from 0 to 13 dBm. As feed power increased Shapiro steps emerge. The modulations of the critical current with dc magnetic field were also measured. Whereas the Shapiro steps were clearly observed, the modulation of the critical current with dc magnetic field differs from the ideal Fraunhofer pattern: the maxima of the critical current occurred at a nonzero applied field, hysteresis was observed when the field was swept in forward and reverse directions. There was some variation in critical current for repeated cooling. We attribute this to trapped magnetic flux in the superconducting films,[14,15] as the cryostat was not well shielded and also due to misalignment of the junction in the magnetic field. The possibility of having defects in barrier is less probable since $I_C$ and $R_N$ scale with the junction area.

Results for MgB$_2$/native oxide/MgB$_2$ tunnel junctions are shown in Figure 3. Fig. 3a depicts the conductance vs. bias voltage at selected temperatures for such a junction with $R_N$ =6.2 kΩ. The sum of the superconducting energy gaps of the top and bottom MgB$_2$ electrodes at 4.2 K is 4.8 mV. Assuming contributions only from π-band, the gap values of 2.4 mV are consistent with previous results.[16] The junction exhibits superconducting energy gaps at temperatures above 30 K. However the presence of sub-gap characteristics suggests high leakage barriers and also no supercurrent was observed. The temperature dependence of the sum-gap (Fig. 3b) is well fitted using expression[2] $\Delta(T)=\Delta(0)*[1- (T/T_C)^p]^{1/2}$ for $\Delta(0)=2.4$ mV, $T_C$=36 K, and p=2.4. Similar results were reported in a recent experiment by Singh *et al.*[8] ($\Delta(0)$=2.2 mV, Tc=36.5 K and p=2). Finally, the conductance versus voltage measurements at higher voltages (>100 mV) were used to evaluate barrier height and thickness using Simmons's model.[17] We found that barrier height ranges from 0.1 to 0.15 eV and thickness from 4 to 5 nm. The barrier height is much smaller than the one for deposited MgO barriers (~ 0.9 eV) showing highly "defective" nature of the barrier, most likely due to the composite nature (such as $MgB_xO_y$) which might form inelastic tunneling trap states inside the barrier. Such a phenomenon has been observed by the J. Read *et al.*.[18]

In comparison with earlier studies, the above results agree with Singh *et al.*,[8] where barrier height ranges from 0.11 to 0.33 eV and thickness from 4.1 to 5.4 nm for $MgB_2$/native oxide/$MgB_2$ junctions, while differ with Schneider *et al.*[7] for $MgB_2$/native oxide/In where they found the barrier height and thickness to be 1.6 eV and 1.5 nm. These results show that the oxidation process is complicated, since the barrier behavior is determined by surface stoichiometry of $MgB_2$ film which depends on the film deposition process.

In conclusion, we have prepared *all* $MgB_2$ junctions with barriers made of deposited MgO as well as native oxide. The MgO based junctions showed Josephson tunneling characteristics up to 20 K although $T_C$ values were well above that temperature. We attribute this to the thermal noise, and also possibly due to the top electrode degradation during the fabrication process. With optimized fabrication process and measurements in a shielded cryostat, the Josephson tunneling characteristics may likely be measured even at higher temperatures. Junctions with native oxide barriers showed thick and poor oxide layer and supercurrent were not observed. These latter junctions also revealed sub-gap structures. For the fabrication of *all* $MgB_2$ based Josephson Junctions devices, our results demonstrate that the junction oxidation process is less reliable compared to growing an artificial barrier. Moreover, with the performance reported here, the *all* $MgB_2$ Josephson Junctions with MgO barriers may find wide-ranging application especially in medical field among others.

We acknowledge the Office of Naval Research funding N00014-06-01-0235 for providing financial support.

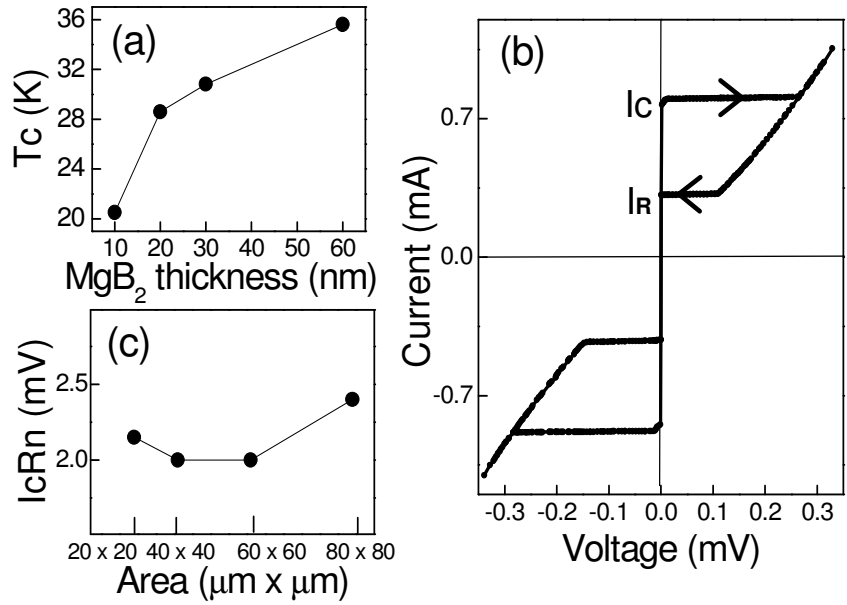

**Figure 1 (a)** Variation of superconducting transition temperature with MgB$_2$ film thickness. **(b)** Typical I-V characteristic of an MgB$_2$/MgO/MgB$_2$ junction at 4.2 K with 2.4 nm MgO, showing dc Josephson Effect. The hysteresis suggests underdumped junctions with Stewart-McCumber parameter $(4I_C/\pi\, I_R)^2$ equal to 12. **(c)** $I_C R_N$ product vs. junction area. Poor dependence on junction area indicates that the junctions were leak-current free.

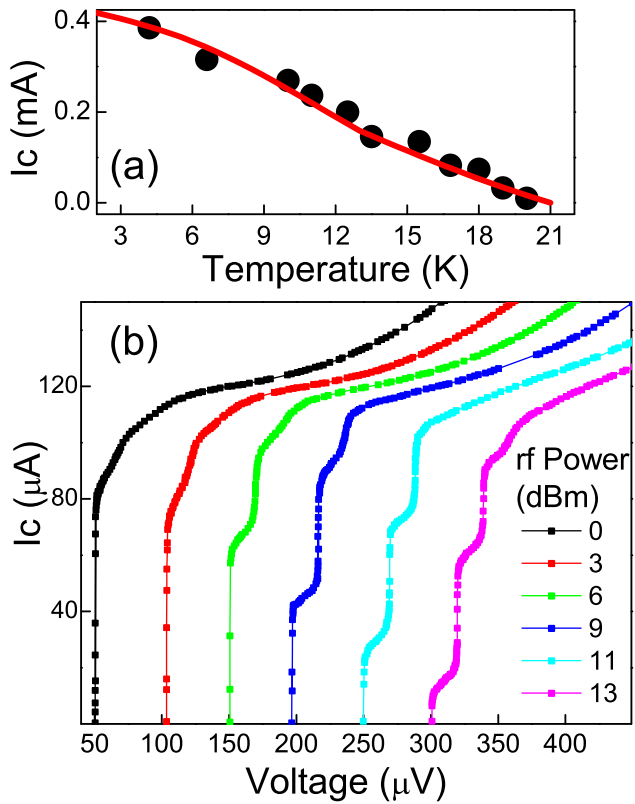

**Figure 2 (a)** The $I_C$ dependence of temperature, the filed circles are experimental data and the solid line is fit with expression $I_C \approx (\Delta(t)/eR_N)\tanh(\Delta(t)/2k_BT)$, assuming two gaps superconductor. **(b)** I-V characteristics under irradiation of 9.3 GHz at 4.2 K measured for different rf power. The junction size was 60×60 µm$^2$. For clarity the plots were shifted 50 µV along the horizontal. The Shapiro steps appeared at the expected voltage values of n times V= (h/2e)*f = 20.7 µV with 2e/h = 483.6 MHz/V.

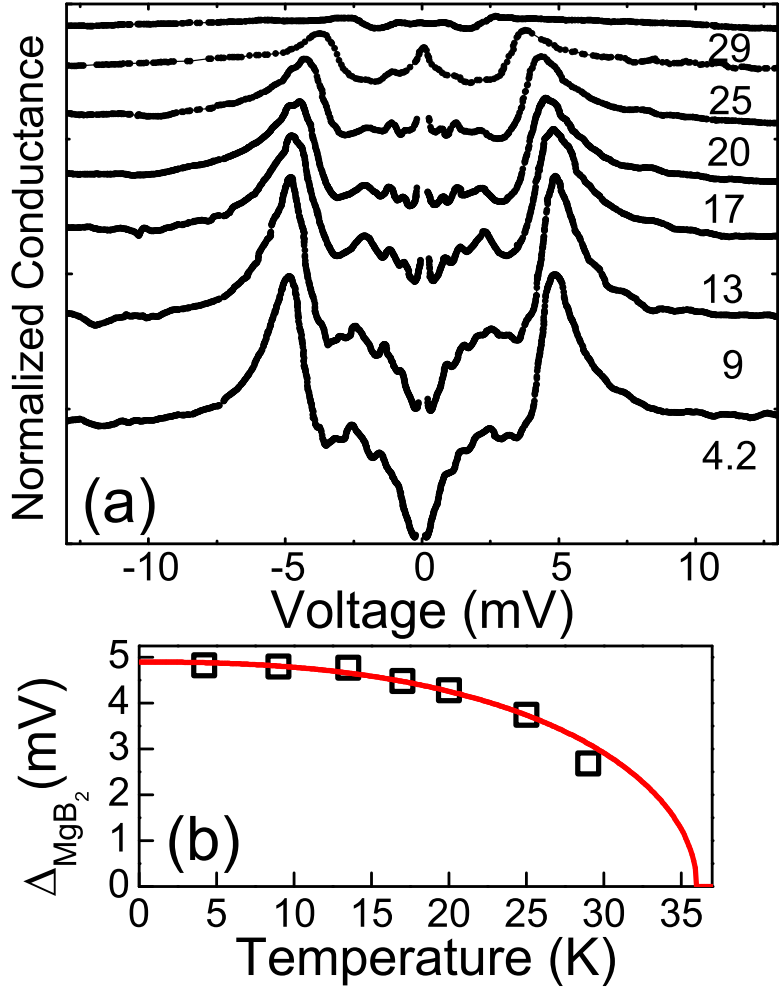

**Figure 3 (a)** Normalized conductance vs. voltage at selected temperatures for an MgB$_2$/native oxide/MgB$_2$ junction. The junction size was 100×150 µm$^2$. The sub-gap characteristics suggest leakage barrier. **(b)** The superconducting energy gap is well fitted using expression $\Delta(T)=\Delta(0)*[1- (T/T_C)^p]^{1/2}$ for $\Delta(0)$=2.4 mV, Tc=36 K, and p=2.4.